\documentclass[10pt]{article}
\usepackage{amsmath,amssymb,amsthm}
\usepackage{graphicx}

\newtheorem{definition}{Definition}
\newtheorem{proposition}{Proposition}

\newcommand{\be}{\begin{equation}}
\newcommand{\ee}{\end{equation}}
\newcommand{\bea}{\begin{eqnarray}}
\newcommand{\eea}{\end{eqnarray}}
\newcommand{\ba}{\begin{array}}
\newcommand{\ea}{\end{array}}

\newcommand{\bp}{\begin{proposition}}
\newcommand{\ep}{\end{proposition}}
\newcommand{\bd}{\begin{definition}}
\newcommand{\ed}{\end{definition}}

\font\mediumroman=cmr10 at 8pt

\begin{document}

\title{A Geometrical Representation of Entanglement as Internal Constraint}
\author{Diederik Aerts, Ellie D'Hondt\footnote{Research Assistant of the Fund for Scientific Research -
Flanders (Belgium)} and Bart D'Hooghe\footnote{ Postdoctoral
Fellow of the Fund for Scientific Research - Flanders (Belgium)}}
\date{}
\maketitle

\centerline{Centrum Leo Apostel (CLEA), Vrije Universiteit Brussel} %
\centerline{Krijgskundestraat 33, B1160 Brussels, Belgium} %
\centerline{diraerts,eldhondt,bdhooghe@vub.ac.be}

\begin{abstract}
\noindent We study a system of two entangled spin 1/2, were the spin's are represented by a sphere model
developed within the hidden measurement approach which is a generalization of the Bloch sphere
representation, such that also the measurements are represented. We show how an arbitrary tensor
product state can be described in a
complete way by a specific internal constraint between the ray or density
states of the two spin 1/2. We derive a geometrical view of entanglement as
a `rotation' and `stretching' of the sphere representing the states of the
second particle as measurements are performed on the first particle.
In the case of the singlet state entanglement can be
represented by a real physical constraint, namely by means of a rigid rod.

\end{abstract}


\section{Introduction}

Within the hidden measurement approach to quantum mechanics \cite
{Aerts1986,Aerts1987} entanglement has been studied for a system consisting
of two entangled spin $\frac 12$ particles in the singlet state \cite
{Aerts1991,aerts1991hpa}. In such a case, typical EPR correlations are
encountered, meaning that if one of both spins collapses in a certain
direction under the influence of a measurement, then the other spin
collapses in the opposite direction. In \cite{coecke1998,coecke2000} these
results were generalized to give a description of entanglement as a hidden
correlation between the proper states of the individual subsystems. Our aim
is to elaborate on these results, more specifically, we want to develop a
geometrical representation of entanglement by means of an internal
constraint between the states of the spin 1/2 particles, represented on the
sphere, for an arbitrary tensor product state that is not necessarily the
singlet state. We do this by introducing constraint functions, which
describe the behavior of the state of one of the spins if measurements are
executed on the other spin. In \cite{aerts1991hpa} the internal constraint
was given a real physical classical mechanics representation, namely by
means of a rigid rod. An interesting question remained whether it is also
possible for non singlet states to invent a similar mechanistic device.

For the individual spin 1/2 entities we use a sphere model
representation developed within the hidden measurement approach to quantum
mechanics \cite
{Aerts1986,Aerts1987,Aerts1991,aerts94a,aerts94b,aerts1997,aerts1997b}%
, which is a generalization of the Bloch or Pauli representation, such that
also the measurements are represented. We identify a parameter $r\in \left[
0,1\right] $, arising from the Schmidt diagonal decomposition, that is a
measure of the amount of entanglement, such that for $r=0$ the system is in
the singlet state with maximal entanglement (and we recover previous
results), and for $r=1$ the system is in a pure product state. For
intermediate values of $r$ we encounter new situations in which entanglement
is expressed by a rotation and distortion of the sphere, representing the
state of the single spin 1/2 entities.

Concerning measurements and their effect on one of the spins in an
entangled state when executed on the other spin, we consider on the one hand
a measurement on a pure state followed by a collapse of the state, as
prescribed by Von Neumann's formula, and on the other hand a measurement on
a mixed state resulting in a new mixed state, as prescribed by Luder's
formula. We will show that an arbitrary collapse measurement on one spin
provokes a rotation and a stretching on the other spin, which can be
described in detail by means of the sphere model, and an arbitrary
measurement on one of the two spins in a density state does not provoke any
change in the partial trace density matrix of the other spin, i.e., the
spins behave as separated entities for such measurements.

\section{The Sphere Model}

The sphere model is a generalization of the Bloch sphere representation,
such that also the measurements as well as a parameter for non-determinism
can be represented \cite{aerts1997b}. In this model, a spin 1/2 state ${%
|\psi \rangle }=\left( \cos {\frac \theta 2}e^{\frac{-i\phi }2},\sin {\frac
\theta 2}e^{\frac{i\phi }2}\right) $ is represented by the point $u(1,\theta
,\phi )=(\sin \theta \cos \phi ,\sin \theta \sin \phi ,\cos \theta )$ on the
surface of a 3-dimensional unit sphere, often called Bloch or Poincar\'{e}
sphere. All points of the Bloch sphere represent states of the spin, such
that points on the surface correspond to pure states, while interior points
correspond to density states. This is because an arbitrary point $u(r,\theta
,\phi )$, $r\in [0,1],\theta \in [0,\pi ],\phi \in [0,2\pi ]$, of the Bloch
sphere can in general be written as a convex linear combination $u(r,\theta ,\phi )=ru(1,\theta ,\phi
)+(1-r)u(0,\theta ,\phi )$ from which follows the corresponding density state 
\begin{equation}
D(r,\theta ,\phi )=rD(1,\theta ,\phi )+(1-r)D(0,\theta ,\phi )={\frac 12}%
\left( 
\begin{array}{cc}
1+r\cos \theta  & r\sin \theta e^{-i\phi } \\ 
r\sin \theta e^{i\phi } & 1-r\cos \theta \label{gendensitymatrix}
\end{array}
\right) 
\end{equation}
In this expression $D(1,\theta ,\phi )={|\psi \rangle }{\langle \psi |}$ is
the usual density state representation of a pure state, while $D(0,\theta
,\phi )$ is the density matrix representing the center of the sphere (the singlet state). Next to this,
the sphere model allows a representation of \emph{measurements}. Without loss of generality we can
demonstrate the effect of such a measurement by considering states that are
on the straight line connecting the North pole $D(1,0,\phi )={|0\rangle }{%
\langle }0|$ and the South pole $D(1,\pi ,\phi )={|1\rangle }{\langle }1|$
of the sphere (we use the convention that ${|0\rangle }$ corresponds to spin
up or $(1,0)$, while ${|1\rangle }$ corresponds to spin down or $(0,1)$). In
this case, the spin is in density state $D(r,0,0)$. After a measurement of
the spin in the direction $u(1,\theta ,\phi )$, the density state of the
spin becomes (by means of Luder's Formula) 
\begin{equation}
D=P(\theta ,\phi )D(r,0,0)P(\theta ,\phi )+(1-P(\theta ,\phi
))D(r,0,0)(1-P(\theta ,\phi ))  \label{luder}
\end{equation}
where $P(\theta ,\phi )$ is the projector on the ray state $|\theta \phi
\rangle $, and hence equals $D(1,\theta ,\phi )$. For $\theta \in [0,{\frac
\pi 2}]$, this results in the density matrix 
\begin{equation}
D={\frac 12}\left( 
\begin{array}{cc}
1+r^{\prime }\cos \theta  & r^{\prime }\sin \theta e^{-i\phi } \\ 
r^{\prime }\sin \theta e^{i\phi } & 1-r^{\prime }\cos \theta \label%
{noncollapse}
\end{array}
\right) =D(r^{\prime },\theta ,\phi )
\end{equation}
where $r^{\prime }=r\cos \theta $. A similar expression $D(r^{\prime
},\theta ^{\prime },\phi ^{\prime })$, with $r^{\prime }=r\cos \theta
^{\prime }$, $\theta ^{\prime }=\pi -\theta $ and $\phi ^{\prime }=\phi +\pi 
$, is obtained for $\theta \in [{\frac \pi 2},\pi ]$. If we consider the
sphere we can see easily that in both cases the point $u(r,0,0)$ is
transformed into the point 
\begin{equation}
(u(r,0,0)\cdot u(1,\theta ,\phi ))u(1,\theta ,\phi )
\end{equation}
This means that we have identified a very simple mechanics to describe the
quantum measurement effect on a mixed state in our sphere model. The effect
is just an ordinary orthogonal projection on the direction of the spin
measurement of the point that represents the density state of the spin in
the sphere model, as represented in Fig. 1.

\vskip 0.5 cm
\hskip 3 cm \includegraphics{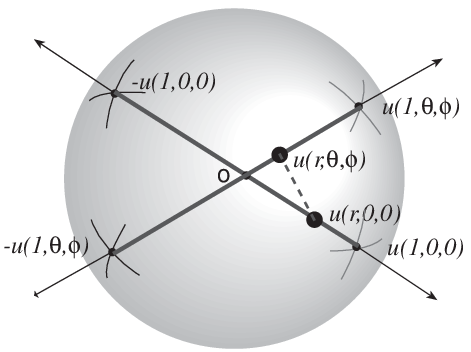}

\vskip 0.3 cm
\hskip 2.2 cm \begin{minipage}{8 cm}
\mediumroman \baselineskip 9 pt Figure 1: Effect of the measurement on a single spin $\frac 12$.
\end{minipage}
\vskip 0.4 cm
\noindent 
In general, suppose that we have a spin state represented by the point $%
u(s,\alpha ,\beta )$ and we perform a spin measurement in direction $(\theta
,\phi )$. If we denote the orthogonal projection on the direction $(\theta
,\phi )$ by $E(\theta ,\phi )$, the new state after this measurement is
given by 
\begin{equation}
\left\{ 
\begin{array}{l}
E(\theta ,\phi )u(s,\alpha ,\beta )=u(s\cos \theta ,\theta ,\phi )\ \mathrm{%
if}\ |\alpha -\theta |\in [0,{\frac \pi 2}] \\ 
E(\theta ,\phi )u(s,\alpha ,\beta )=u(s\cos (\pi -\theta ),\pi -\theta ,\phi
+\pi )\ \mathrm{if}\ |\alpha -\theta |\in [{\frac \pi 2},\pi ]
\end{array}
\right. 
\end{equation}
It is possible to give a nice geometrical presentation of how the spin state
changes under the influence of measurements in different directions, as shown in Fig. 2.

\vskip 0.2 cm
\hskip 2.9 cm \includegraphics{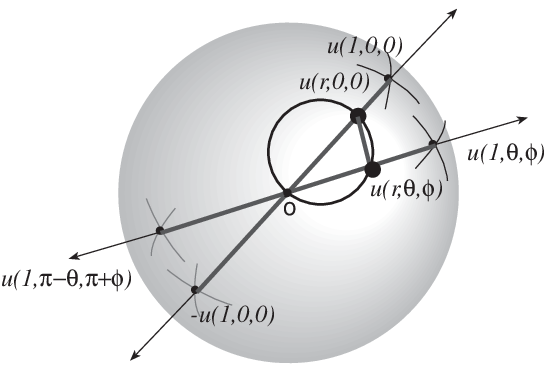}

\vskip 0.3 cm
\hskip 1.9 cm \begin{minipage}{8 cm}
\mediumroman \baselineskip 9 pt Figure 2: A geometrical presentation of how the spin state changes
under the influence of measurements in different directions.
\end{minipage}
\vskip 0.4 cm
\noindent More concretely, consider a little sphere with as North pole the
point $u(s,\alpha ,\beta )$, the point that represents the spin state, and
as South pole the center of the big sphere of the model. The spin state is
transformed to the point of intersection between this little sphere and the
direction of the measurement performed. Hence the points of the little
sphere are those points representing the states where the spin state can be
transformed to, under arbitrary angles of measurement.

\section{Constraint Functions} \label{section3}

A system of two entangled spin ${\frac 12}$ is described by means of an
arbitrary unit vector ${|\psi \rangle }\in {\mathbb C}_1^2\otimes {\mathbb C}%
_2^2$, in which ${\mathbb C}_1^2$ and ${\mathbb C}_2^2$ are two copies of ${%
\mathbb C}^2$, which we label with indices $1$ and $2$ with the sole purpose
of identifying them. The vector ${|\psi \rangle }$ can always be written as
the following linear combination 
${|\psi \rangle }=\sum_{ij}\lambda _{ij}|e_1^i{\rangle }\otimes |e_2^j{%
\rangle }$
where $\lambda _{ij}\in {\mathbb C}$, and $\{|e_1^i{\rangle }\}$ and $%
\{|e_2^j{\rangle }\}$ are bases of ${\mathbb C}_1^2$ and ${\mathbb C}_2^2$
respectively. When we carry out a collapse measurement on the first spin,
Von Neumann's formula describes how it collapses into a spin state described
by the unit vector $|x_1{\rangle }\in {\mathbb C}_1^2$, thus transforming
the entangled state ${|\psi \rangle }$ into 
$(P_{|x_1{\rangle }}\otimes I)({|\psi \rangle })$
where $P_{|x_1{\rangle }}$ is the orthogonal projector on $|x_1{\rangle }$
in ${\mathbb C}_1^2$, and $I$ is the unit operator in ${\mathbb C}_2^2$. The
result is that the entangled spins end up in the following product state 
$|x_1{\rangle }\otimes \sum_{ij}\lambda _{ij}\langle x_1,e_1^i\rangle |e_2^j{%
\rangle }$.
This means that as a consequence of the measurement on the first spin,
collapsing its state to $|x_1{\rangle }$, the second spin collapses to the
state $\sum_{ij}\lambda _{ij}\langle x_1,e_1^i\rangle |e_2^j{\rangle }$.

In an analogous way we can show that if a measurement is performed on the
second spin, resulting in a collapse to the state $x_2\in {\mathbb 
C}_2^2$, the state of the first spin becomes 
$\sum_{ij}\lambda _{ij}\langle x_2,e_2^j\rangle |e_1^i{\rangle }$.
Because of this, we arrive at the following definition.

\begin{definition}[Constraint Functions]
We define the constraint functions $F_{12}(\psi )$ and $F_{21}(\psi )$
related to $\psi $ in the following way 
\begin{eqnarray}
F_{12}(\psi ) &:&{\mathbb C}_1^2\rightarrow {\mathbb C}_2^2:|x_1{\rangle }%
\mapsto \sum_{ij}\lambda _{ij}\langle x_1,e_1^i\rangle |e_2^j{\rangle } \\
F_{21}(\psi ) &:&{\mathbb C}_2^2\rightarrow {\mathbb C}_1^2:|x_2{\rangle }%
\mapsto \sum_{ij}\lambda _{ij}\langle x_2,e_2^j\rangle |e_1^i{\rangle }
\end{eqnarray}
\end{definition}

\noindent In other words, the constraint functions map the state where one
of the spins collapses to by a measurement to the state that the other spin
collapses to under influence of the entanglement correlation. A detailed
study of the constraint functions can give us a complete picture of how the
entanglement correlation works as an internal constraint. Before we arrive
at this complete picture, however, we give some properties of the constraint
functions.

One can show that the following properties hold for the constraint functions
and the relation between the two constraint functions $F_{12}(\psi )$ and $%
F_{21}(\psi )$:

\begin{proposition}
The constraint functions are canonically defined.
\end{proposition}

\begin{proposition}
The constraint functions are conjugate linear.
\end{proposition}

\begin{proposition}
$\left\{ 
\begin{array}{l}
D_1(\psi )\equiv tr_{{\mathbb C}_1^2}{|\psi \rangle }{\langle }\psi
|=F_{21}(\psi )\circ F_{12}(\psi ) \\ 
D_2(\psi )\equiv tr_{{\mathbb C}_2^2}{|\psi \rangle }{\langle }\psi
|=F_{12}(\psi )\circ F_{21}(\psi )
\end{array}
\right. $ or in other words, $F_{21}(\psi )\circ F_{12}(\psi )$ equals $%
D_1(\psi ),$ i.e., the partial trace density matrix over ${\mathbb C}_2^2$
and $F_{12}(\psi )\circ F_{21}(\psi )$ equals $D_2(\psi ),$ i.e., the
partial trace density matrix over ${\mathbb C}_1^2$.
\end{proposition}

\begin{proposition}
For $|x_1{\rangle }\in {\mathbb C}_1^2$ and $|x_2{\rangle }\in {\mathbb C}%
_2^2$ we have 
\begin{equation}
\langle F_{12}(\psi )(|x_1{\rangle }),x_2\rangle =\langle x_1,F_{21}(\psi
)(|x_2{\rangle })\rangle ^{*}  \label{eq:constraintrelated}
\end{equation}
\end{proposition}
\noindent
To derive a complete view of how entanglement works as an internal
constraint for a 2-particle system, we now work out the relation between the
Schmidt diagonal form (e.g. \cite{nielsen2000}) and the constraint
functions. We begin by choosing the base $|x_1^1{\rangle }=(\cos {\frac
\theta 2}e^{-i{\frac \phi 2}},\sin {\frac \theta 2}e^{\frac{i\phi }2}),$ $%
|x_1^2{\rangle }=(-i\sin {\frac \theta 2}e^{-i{\frac \phi 2}},i\cos {\frac
\theta 2}e^{\frac{i\phi }2})$ in ${\mathbb C}_1^2.$ With respect to this
basis, expression (\ref{gendensitymatrix}) for a general density matrix
becomes 
\begin{equation}
D_1(\psi )={\frac 12}\left( 
\begin{array}{cc}
1+r & 0 \\ 
0 & 1-r
\end{array}
\right)
\end{equation}
One can choose a basis $\left\{ |x_2^1{\rangle ,}|x_2^2{\rangle }\right\} $
in ${\mathbb C}_2^2$ given by 
\begin{equation}
|x_2^1{\rangle }={\frac{\sqrt{2}}{\sqrt{1+r}}}F_{12}(\psi )(|x_1^1{\rangle }%
),\ |x_2^2{\rangle }={\frac{\sqrt{2}}{\sqrt{1-r}}}F_{12}(\psi )(|x_1^2{%
\rangle })  \label{x22}
\end{equation}
One can show that $\Vert x_2^1\Vert ^2=1=$ $\Vert x_2^2\Vert ^2$ and 
\begin{equation}
D_2(\psi )(|x_2^1{\rangle })={\frac{1+r}2}|x_2^1{\rangle ,\ }D_2(\psi
)(|x_2^2{\rangle })={\frac{1-r}2}|x_2^2{\rangle }
\end{equation}
Hence $|x_2^1{\rangle }$ and $|x_2^2{\rangle }$ are normalized eigenvectors
of $D_2(\psi )$ with eigenvalues ${\frac{1+r}2}$ and ${\frac{1-r}2}$
respectively. Therefore, with respect to the basis $\{|x_2^1{\rangle },|x_2^2%
{\rangle }\}$, $D_2(\psi )$ is expressed as 
\begin{equation}
D_2(\psi )={\frac 12}\left( 
\begin{array}{cc}
1+r & 0 \\ 
0 & 1-r
\end{array}
\right)
\end{equation}
Finally, let us find the expression for $\psi $ with respect to the basis $%
\{|x_1^1{\rangle }\otimes |x_2^1{\rangle },|x_1^1{\rangle }\otimes |x_2^2{%
\rangle },|x_1^2{\rangle }\otimes |x_2^1{\rangle },|x_1^2{\rangle }\otimes
|x_2^2{\rangle }\}$ of ${\mathbb C}_1^2\otimes {\mathbb C}_2^2$. In general,
this expression is of the form 
$\psi =a|x_1^1{\rangle }\otimes |x_2^1{\rangle }+b|x_1^1{\rangle }\otimes
|x_2^2{\rangle }+c|x_1^2{\rangle }\otimes |x_2^1{\rangle }+d|x_1^2{\rangle }%
\otimes |x_2^2{\rangle }$.
However, since 
\begin{eqnarray}
F_{12}(\psi )(|x_1^1{\rangle }) &=&a|x_2^1{\rangle }+b|x_2^2{\rangle }={%
\frac{\sqrt{1+r}}{\sqrt{2}}}|x_2^1{\rangle } \\
F_{12}(\psi )(|x_1^2{\rangle }) &=&c|x_2^1{\rangle }+d|x_2^2{\rangle }={%
\frac{\sqrt{1-r}}{\sqrt{2}}}|x_2^2{\rangle }
\end{eqnarray}
we obtain 
\begin{equation}
a={\frac{\sqrt{1+r}}{\sqrt{2}},\ }b=0,\ c=0,\ d={\frac{\sqrt{1-r}}{\sqrt{2}}}
\end{equation}
Thus, the Schmidt diagonal form of ${|\psi \rangle }$ is given by 
\begin{equation}
{|\psi \rangle }={\frac{\sqrt{1+r}}{\sqrt{2}}}|x_1^1{\rangle }\otimes |x_2^1{%
\rangle }+{\frac{\sqrt{1-r}}{\sqrt{2}}}|x_1^2{\rangle }\otimes |x_2^2{%
\rangle }  \label{schmidt}
\end{equation}

\section{Measurements\label{section4}}

With all the above we can now concentrate on the role of measurements. More
particularly, we analyze how a measurement, carried out on one subentity of
an entangled system, affects the state of the other subentity of which the
entangled system is composed. We discuss both the effect of a measurement on
density states as described by Luder's formula and the effect of a collapse
measurement on a pure state as described by Von Neumann's formula.

To describe the effect of a measurement on the density state of a subsystem
we use Luder's formula (\ref{luder}), where in this case the initial density
state is calculated from the Schmidt diagonal form (\ref{schmidt}) derived
above. Choosing bases $\left\{ x_1^1=\left( 1,0\right) ,x_1^2=\left(
0,1\right) \right\} $ in ${\mathbb C_1^2}$ and $\left\{ x_2^1=\left(
1,0\right) ,x_2^2=\left( 0,1\right) \right\} $ in $\mathbb{C}_2^2$, one can
calculate the density state $D\left( \psi \right) =\left| \psi \right\rangle
\left\langle \psi \right| $ corresponding with the pure state $\left| \psi
\right\rangle .$ After the measurement this state has changed into the
density state $D^{\prime }\left( \psi \right) $ given by Luder's formula: 
$D^{\prime }\left( \psi \right) =\left( P(\theta ,\phi )\otimes \mathbf{1}%
\right) D\left( \psi \right) \left( P(\theta ,\phi )\otimes \mathbf{1}%
\right) +\left( \left( \mathbf{1}-P(\theta ,\phi )\right) \otimes \mathbf{1}%
\right) D\left( \psi \right) \left( \left( \mathbf{1}-P(\theta ,\phi
)\right) \otimes \mathbf{1}\right)$,
from which we can calculate $D_1(\psi ),$ i.e., the partial trace density
matrix to ${\mathbb C}_1^2$, obtaining 
\begin{equation}
D_1(\psi )={\frac 12}\left( 
\begin{array}{cc}
1+r\cos ^2\theta & r\sin \theta \cos \theta e^{-i\phi } \\ 
r\sin \theta \cos \theta e^{i\phi } & 1-r\cos ^2\theta \label{d1}
\end{array}
\right)
\end{equation}
This is the same density matrix as we found in expression (\ref{noncollapse}%
), i.e. after carrying out a measurement on a single spin $\frac 12$ in a
density state. On the other hand, if we calculate $D_2(\psi ),$ i.e., the
partial trace density matrix to ${\mathbb C}_2^2$, we find: 
\begin{equation}
D_2(\psi )=\frac 12\left( 
\begin{array}{ll}
1+r & 0 \\ 
0 & 1-r
\end{array}
\right)  \label{d2}
\end{equation}
which is independent of $\left( \theta ,\phi \right) .$ From expressions (%
\ref{d1}) and (\ref{d2}), one derives that a measurement prescribed by
Luder's formula on one spin does not provoke any change in the partial trace
density matrix of the other spin: in other words, the spins behave as
separated entities for such measurements.

Let us now study what happens when a collapse measurement is performed on
one of the subsystems in the entangled system. Since the constraint
functions describe exactly this, studying collapse measurements means
studying the constraint function, more specifically how they map points of
the sphere (i.e., quantum states) onto one another. As a point of departure,
we choose $\psi $ and bases as in the above, where both bases are connected
through equation (\ref{x22}). From these equations, we first observe that
within the sphere model, they imply that the north (south) pole of the first
sphere is mapped onto the north (south) pole of the second sphere. Next, it
follows immediately that $F_{12}(\psi )$ does not conserve the norm. Indeed,
the norm of $F_{12}(\psi )(|x{\rangle })$ for an arbitrary vector $|x{%
\rangle }=x(\theta ,\phi )$ is as follows: 
\begin{equation}
\Vert F_{12}(\psi )(|x{\rangle })\Vert ^2=\frac{1+r}2\cos ^2{\frac \theta 2}+%
\frac{1-r}2\sin ^2{\frac \theta 2}=\frac 12\left( 1+r\cos \theta \right)
\label{norm}
\end{equation}
If we consider for a moment the angle $\theta $ as a variable, we see that
the square of the norm varies between $\frac{1+r}2$ and $\frac{1-r}2$, for
the north ($\theta =0$) and the south ($\theta =\pi $) pole of the sphere
respectively. Actually, this is where the factors $\sqrt{\frac 2{1+r}}$ and $%
\sqrt{\frac 2{1-r}}$ in the original definition of $|x_2^1{\rangle }$ and $%
|x_2^2{\rangle }$ in equation (\ref{x22}) come from. Not
only the norm, but also orthogonality is in general not conserved by $%
F_{12}(\psi )$. For example, using the conjugate linearity of the constraint
functions, we find that the two orthonormal vectors $|\psi _u{\rangle }=\psi
(\theta ,\phi )$ and $|\psi _{-u}{\rangle }=\psi (\pi -\theta ,\phi +\pi )$
are mapped to 
\begin{eqnarray}
F_{12}(\psi )(\psi _u) &=&\sqrt{\frac{1+r}2}\cos {\frac \theta 2}\ e^{i{%
\frac \phi 2}}\ x_2^1+\sqrt{\frac{1-r}2}\sin {\frac \theta 2}\ e^{-i{\frac
\phi 2}}\ x_2^2 \\
F_{12}(\psi )(\psi _{-u}) &=&\sqrt{\frac{1+r}2}i\sin {\frac \theta 2}\ e^{i{%
\frac \phi 2}}\ x_2^1-i\sqrt{\frac{1-r}2}\cos {\frac \theta 2}\ e^{-i{\frac
\phi 2}}\ x_2^2
\end{eqnarray}
For $0\neq \theta \neq \pi $ orthogonality is
conserved if
$\langle F_{12}(\psi )(|\psi _u{\rangle }),F_{12}(\psi )(|\psi _{-u}{\rangle }%
)\rangle =0$ with means that $r=0$.
Translated on the sphere this means that diametrical opposite points are
mapped to diametrical opposite points only in the special case $r=0$ (except
for the north and south pole, which are always mapped onto the north and
south pole of the second sphere). In other words, orthogonality only is
generally conserved for the singlet state. While the norm and orthogonality are in general not conserved,
we can look at the \emph{normalized} image corresponding to a state $|x{\rangle 
}=x(\theta _1,\phi _1)$. In other words, we would like to know where the
state 
\begin{equation}
|y{\rangle }=y(\theta _2,\phi _2)={\frac 1{\Vert F_{12}(\psi )(|x{\rangle }%
)\Vert }}F_{12}(\psi )(|x{\rangle })
\end{equation}
lies on the sphere. Therefore, we compare corresponding inproducts on both
spheres, and we obtain that

\begin{equation}
\langle y,x_2^1\rangle =\sqrt{\frac{1+r}{1+r\cos \theta _1}}\cdot \langle
x,x_1^1\rangle ^{*}  \label{inproducts}
\end{equation}
Again, we see that only for the singlet state inproducts are equal (and
consequently, antipodal points on the sphere are mapped to antipodal points,
as mentioned above). An interesting case is to look at the image of the
equator, or in other words the points for which $\theta _1={\frac \pi 2}$.
In this case 
\begin{equation}
\langle y,x_2^1\rangle =\sqrt{1+r}\cdot \langle x,x_1^1\rangle ^{*}=\sqrt{1+r%
}{\frac 1{\sqrt{2}}}e^{-i{\frac{\phi _1}2}}
\end{equation}
Translating this to the sphere model, using the following formula which
expresses the relation between the inproduct of two vectors in ${\mathbb C}%
_2^2$ and the scalar product of the corresponding points in the sphere
representation: 
\begin{equation}
{\frac{{1+\psi (\theta ^{\prime },\phi ^{\prime })\cdot \psi (\theta ,\phi )}%
}2}=|\langle \psi (\theta ^{\prime },\phi ^{\prime }),\psi (\theta ,\phi
)\rangle |^2  \label{productrelation}
\end{equation}
applied to $|y{\rangle }$ and $|x_2^1{\rangle }$, we obtain 
\begin{equation}
{\frac{{1+y(\theta _2,\phi _2)\cdot x_2^1(\theta ,\phi )}}2}=\frac{1+r}2
\end{equation}
and as a consequence: 
${y(\theta _2,\phi _2)}\cdot x_2^1(\theta ,\phi )=r$.
This means that on the sphere, the elements of the equator are mapped onto a
cone that makes an angle $\beta $ with the north south axis of the second
sphere, such that $\cos \beta =r.$ Once more, only for $r=0$ this is again
an equator, hence conserving the angle between the elements of the equator
and the north pole. For $r\in \left] 0,1\right[ $ we obtain a cone with an
angle $0<\beta <{\frac \pi 2}$, which means that the equator has `raised' to
the north. For $r$ approaching $1$ the sphere is stretched more and more to
the north pole of the second sphere. Remember that in this limit case the
superposition state becomes a product state, and this fits with the fact
that for product states indeed the map $F_{12}(\psi )$ maps the first
element of the product to the second. To see the general scheme we use
equation (\ref{inproducts}), which yields 
\begin{equation}
{y(\theta _2,\phi _2)}\cdot x_2^1(\theta ,\phi )={\frac{r+{\cos }\theta _1}{%
1+r\cos \theta _1}}  \label{endresult}
\end{equation}
From this result it follows that straight lines through the center of the
left sphere are mapped onto straight lines through the point $u\left(
r,0,0\right) $ along the north south axis in the second sphere. This gives a
nice geometrical representation of this `stretching' on the second sphere,
as shown in Fig. 3. Again, this shows that indeed
only for the singlet state antipodal points of the first sphere are mapped
onto antipodal points of the second sphere.

\vskip 0.7 cm
\hskip -1 cm \includegraphics{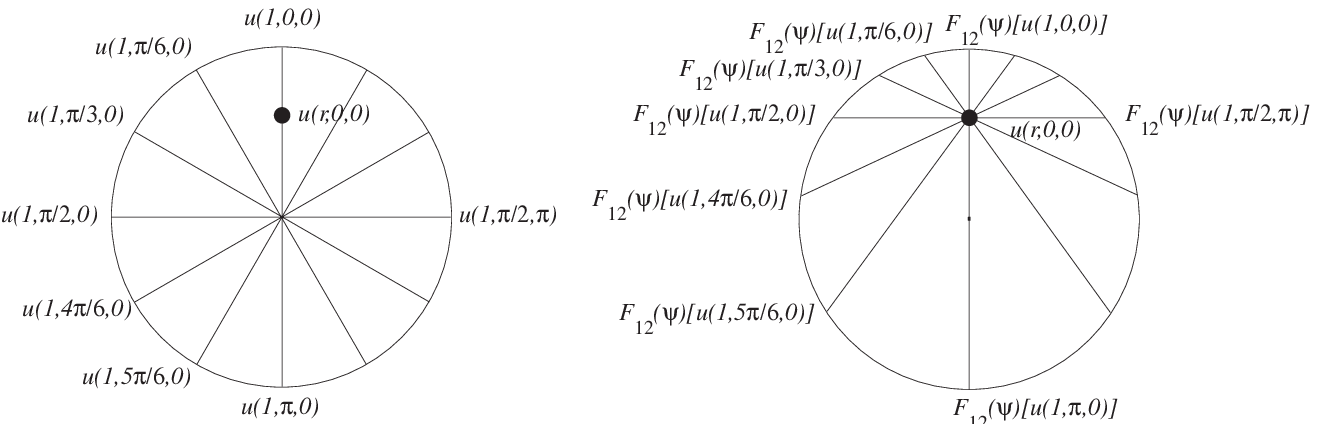}

\vskip 0.3 cm
\hskip 0.4 cm \begin{minipage}{10.5 cm}
\mediumroman \baselineskip 9 pt Figure 3: Straight lines through the center of the left sphere are
mapped onto straight lines through the point $\scriptstyle u\left( r,0,0\right)$ along the
north south axis in the second sphere.
\end{minipage}

\section{Conclusions} \label{section5}
We elaborated a formalism to
model entanglement as an internal constraint. More specifically, we show
that two spin 1/2 particles in a nonproduct state can be
described in a complete way by extracting entanglement into an internal
constraint between the states of the particles. We introduce constraint functions,
 which describe the behavior of the state of one of the spins if measurements are executed on the other
spin. In this way we can substitute the nonproduct state by the states
of the individual particles and the internal constraint function. We make use of the sphere model
representation for the spin's that was developed in Brussels, allowing for an easy
to grasp visual support for the developed formalism. In deriving the effect
measurements on one spin of an entangled state have on the other one, we
differentiated between two types of measurements: measurements, of which the
action on a mixture of states is described by Luder's formula, and collapse
measurements, of which the action is described by Von Neumann's formula. Our
result is that (1) an arbitrary Luder's measurement on one spin in a mixed state
does not provoke any change in the partial trace density matrix of the other
spin, i.e., the spins behave as separated entities for such measurements;
(2) an arbitrary collapse measurement on one spin provokes a rotation and a
stretching on the other spin, which gives a nice geometrical representation
of how entanglement works as an internal constraint. The singlet state appears as a very special case in
which norm and orthogonality are conserved. This makes it easier to understand that for 
the singlet state a real physical apparatus modelling the internal constraint
can be built, namely a rigid rod connecting the two spins. Since for non singlet states norm and
orthogonality are not conserved, and the geometrical representation entails rotation and more
importantly stretching of the sphere, it is not obvious that a simple
machinery (e.g. with a rigid rod) can be constructed in this case.

\end{document}